\documentstyle[amssymb,aps,preprint]{revtex}
\tightenlines

\begin{document}
\title{Improved perturbation theory in the vortex liquids state of type II
superconductors}
\author{Dingping Li\thanks{%
e-mail: lidp@phys.nthu.edu.tw} and Baruch Rosenstein\thanks{%
e-mail: baruch@vortex1.ep.nctu.edu.tw}}
\address{{\it National Center for Theoretical Sciences and} \\
{\it Electrophysics Department, National Chiao Tung University } \\
{\it Hsinchu 30050, Taiwan, R. O. C.}}
\date{\today}
\maketitle

\begin{abstract}
We develop an optimized perturbation theory for the Ginzburg - Landau
description of thermal fluctuations effects in the vortex liquids. Unlike
the high temperature expansion which is asymptotic, the optimized expansion
is convergent. Radius of convergence on the lowest Landau level is $a_{T}=-3$
in 2D and $a_{T}=-5$ in 3D. It allows a systematic calculation of
magnetization and specific heat contributions due to thermal fluctuations of
vortices in strongly type II superconductors to a very high precision. The
results are in good agreement with existing Monte Carlo simulations and
experiments. Limitations of various nonperturbative and phenomenological
approaches are noted. In particular we show that there is no exact
intersection point of the magnetization curves both in 2D and 3D.
\end{abstract}

\vskip 0.5cm 
\flushleft{PACS numbers: 74.60.-w, 74.40.+k,  74.25.Ha,
74.25.Dw}

\newpage

\section{Introduction}

Thermal fluctuations play a much larger role in high $T_{c}$ superconductors
than in the low temperature ones because the Ginzburg parameter $Gi$
characterizing fluctuations is much larger \cite{Blatter}. In addition the
presence of magnetic field and strong anisotropy in superconductors like
BSCCO effectively reduces their dimensionality thereby further enhancing
effects of thermal fluctuations. Under these circumstances the mean field
line separating Abrikosov lattice from ''normal'' phase becomes a phase
transition between vortex lattice and liquid far below the mean field phase
transition line \cite{Nelson,Blatter} clearly seen in both magnetization 
\cite{Zeldov} and specific heat experiments \cite{Schilling}. Between the
mean field transition line and the melting point physical quantities like
the magnetization, conductivity and specific heat depend strongly on
fluctuations. Several experimental observations call for a refined precise
theory. For example a striking feature of magnetization curves intersecting
at the same point $(T^{\ast },H^{\ast })$ was observed in a wide rage of
magnetic fields in both layered (2D or quasi 2D) \cite{Magn2D} materials and
more isotropic ones \cite{Magn3D}. To develop a quantitative theory of these
fluctuations even in the case of the lowest Landau level (LLL) corresponding
to regions of the phase diagram ''close'' to $H_{c2},$ is a very nontrivial
task and several different approaches were developed.

Long time ago Thouless and Ruggeri \cite{Ruggeri,Ruggeri1} proposed a
perturbative expansion around a homogeneous (liquid) state in which all the
''bubble'' diagrams (see Fig. 5) are resummed. Unfortunately they proved
that the series are asymptotic and although first few terms provide accurate
results at very high temperatures, the series become inapplicable for LLL
dimensionless temperature $a_{T}$ $\thicksim $ $(T-T_{mf}(H))/(TH)^{1/2}$
smaller than $2$ in 2D quite far above the melting line (believed to be
located around $a_{T}=-12)$. Generally attempts to extend the theory to
lower temperatures by the Borel transform or Pade extrapolation were not
successful \cite{Moore1}. Several nonperturbative methods have been also
attempted. Originally the RG method was proposed \cite{Nelson} and developed 
\cite{MooreRG} although, since the transition is first order, no solutions
of the RG equations can been found. The set of perturbative ''parquet''
diagrams \cite{Yeo} have been resummed and the large $N$ limit have been
considered \cite{Affleck}. Tesanovic and coworkers developed a method based
on an approximate separation of the two energy scales \cite{Tesanovic} in
both 2D and 3D. The larger contribution (98\%) is the condensation energy,
while the smaller one (2\%) describes motion of the vortices. The theory
explains the intersection of the magnetization curves. This question has
been tackled in 2D by rather phenomenological approach in \cite{Bulaevskii}.
Some Monte Carlo simulations are available \cite{MC,Sasik}.

Meantime experimental precision increased dramatically. New methods like
measurement of magnetization using the Hall probes \cite{Zeldov} were
invented. One can achieve a precision that allows clearly to see a tiny
magnetization jumps of only $0.1Oe$ in BSCCO and a sharp peak in specific
heat in YBCO.

In this paper we apply optimized perturbation theory (OPT) first developed
in field theory \cite{Stevenson,Okopinska,Kleinert} to both the 2D and 3D
LLL model. It allows to obtain a convergent series (rather than asymptotic)
and therefore to calculate magnetization and specific heat of vortex liquids
with definite precision. The precision for various values of the LLL scaled
temperature $a_{T}$ are given in Table 3 and 4. The radius of convergence is 
$a_{T}=-3$ in 2D and $a_{T}=-5$ in 3D. One the basis of this one can make
several definitive qualitative conclusions. The intersection of the
magnetization lines in only approximate not only in 3D (the result already
observed in Monte Carlo simulation \cite{Sasik}), but also in 2D. The theory
by Tesanovic et al \cite{Tesanovic} in 2D describes the physics remarkably
well in high temperatures, but deviates on the 5-10\% precision level at $%
a_{T}=-2$.

The paper is organized as follows. The models are defined and the general
OPT described in section II. The 2D and the 3D calculations are described in
III. Results and comparison with other theories and experiments are given in
section IV. We conclude in section V.

\section{Models and general idea of optimized perturbation theory.}

\subsection{The 2D model}

Our starting point is the Ginzburg-Landau free energy: 
\begin{equation}
F=L_{z}\int d^{2}x\frac{{\hbar }^{2}}{2m_{ab}}\left| D\psi \right|
^{2}+a|\psi |^{2}+\frac{b^{\prime }}{2}|\psi |^{4},
\end{equation}
where ${\bf A}=(By,0)$ describes a nonfluctuating constant magnetic field in
Landau gauge and ${\bf D}\equiv {\bf \nabla }-i\frac{2\pi }{\Phi _{0}}{\bf A,%
}\Phi _{0}\equiv \frac{hc}{e^{\ast }}$. For strongly type II superconductors
like the high $T_{c}$ cuprates ($\kappa \sim 100$) and not too far from $%
H_{c2}$ (this is the range of interest in this paper, for the detailed
discussion of the range of applicability see \cite{Li1}) magnetic field is
homogeneous to a high degree due to superposition from many vortices. For
simplicity we assume $a(T)=\alpha T_{c}(1-t)$, $t\equiv T/T_{c},$ although
this temperature dependence can be easily modified to better describe the
experimental $H_{c2}(T)$.

Throughout most of the paper will use the coherence length $\xi =\sqrt{{%
\hbar }^{2}/\left( 2m_{ab}\alpha T_{c}\right) }$ as a unit of length and $%
\frac{dH_{c2}(T_{c})}{dT}T_{c}=\frac{\Phi _{0}}{2\pi \xi ^{2}}$ as a unit of
magnetic field. After the order parameter field is rescaled as $\psi
^{2}\rightarrow \frac{2\alpha T_{c}}{b^{\prime }}\psi ^{2}$, the
dimensionless free energy (the Boltzmann factor) is: 
\begin{equation}
\frac{F}{T}=\frac{1}{\omega }\int d^{2}x\left[ \frac{1}{2}|D\psi |^{2}-\frac{%
1-t}{2}|\psi |^{2}+\frac{1}{2}|\psi |^{4}\right] ,  \label{energ1}
\end{equation}
where $L_{z}$ is width of the sample. The dimensionless coefficient
describing the strength of fluctuations is 
\begin{equation}
\omega =\sqrt{2Gi}\pi ^{2}t=\frac{m_{ab}b^{\prime }}{2{\hbar }^{2}\alpha }%
t,\;Gi\equiv \frac{1}{2}\left( \frac{32\pi e^{2}\kappa ^{2}\xi ^{2}T_{c}}{%
c^{2}h^{2}L_{z}}\right) ^{2}  \label{omega}
\end{equation}
where $Gi$ is the Ginzburg number in 2D . When $\frac{1-t-b}{12b}<<1$, the
lowest Landau level (LLL) approximation can be used \cite{Li1}. The model
then simplifies due to the LLL constraint, $-\frac{D^{2}}{2}\psi =\frac{b}{2}%
\psi $ to: 
\begin{equation}
f\equiv \frac{F}{T}=\frac{1}{\omega }\int d^{2}x\left[ -\frac{1-t-b}{2}|\psi
|^{2}+\frac{1}{2}|\psi |^{4}\right] .  \label{enlow}
\end{equation}
This reduced model exhibits the LLL scaling. Rescaling again $x\rightarrow x/%
\sqrt{b},y\rightarrow y/\sqrt{b}$ and $|\psi |^{2}\rightarrow |\psi |^{2}%
\sqrt{\frac{b\omega }{4\pi }},$ one obtains 
\begin{equation}
f=\frac{1}{4\pi }\int d^{2}x\left[ a_{T}|\psi |^{2}+\frac{1}{2}|\psi |^{4}%
\right] ,  \label{resca}
\end{equation}
where the 2D LLL reduced temperature 
\begin{equation}
a_{T}\equiv -\sqrt{\frac{4\pi }{b\omega }}\frac{1-t-b}{2}  \label{aT2D}
\end{equation}
is the only parameter in the theory \cite{Thouless,Ruggeri}. In total, we
have done the rescaling 
\begin{equation}
|\psi |^{2}\rightarrow |\psi |^{2}\left( \frac{2\alpha T_{c}}{b^{\prime }}%
\right) \left( \sqrt{\frac{b\omega }{4\pi }}\right) ,x\rightarrow \xi x/%
\sqrt{b},y\rightarrow \xi y/\sqrt{b.}  \label{rescaling}
\end{equation}

We will be interested in thermodynamic properties of the model determined by
partition function $Z=\int D\psi D\overline{\psi }\exp \left( -f\right) $
and will mainly study only the rescaled partition function $%
Z_{r}(a_{T})=\int D\psi _{r}D\overline{\psi }_{r}\exp \left( -f\right) =Z/J,$
where $J$ is a Jacobian. Consequently to obtain, for example, the free
energy density from the\ corresponding quantity in the rescaled model $%
f_{eff}=-\frac{4\pi \log Z_{r}}{V^{^{\prime }}},$ one should use the
following relation: 
\begin{eqnarray}
-\frac{T\log Z}{V} &=&\frac{T}{4\pi }\frac{V^{^{\prime }}}{V}\frac{\left(
-4\pi \log Z_{r}J\right) }{V^{^{\prime }}}  \nonumber \\
&=&\frac{T}{2\pi }\left( \frac{\sqrt{b}}{\xi }\right) ^{2}\log \left( \frac{%
2\alpha T_{c}}{b^{\prime }}\times \sqrt{\frac{b\omega }{4\pi }}\right) +%
\frac{T}{4\pi }\left( \frac{\sqrt{b}}{\xi }\right) ^{2}f_{eff}.
\end{eqnarray}
From now on we work with rescaled quantities only and related them to
measured quantities in section IV.

\subsection{The 3D model}

For 3D, the GL model takes a form 
\begin{equation}
F=\int d^{3}x\frac{{\hbar }^{2}}{2m_{ab}}\left| \left( {\bf \nabla }-\frac{%
ie^{\ast }}{\hbar c}{\bf A}\right) \psi \right| ^{2}+\frac{{\hbar }^{2}}{%
2m_{c}}|\partial _{z}\psi |^{2}+a|\psi |^{2}+\frac{b^{\prime }}{2}|\psi |^{4}
\end{equation}
which can be again rescaled into 
\begin{equation}
f=\frac{F}{T}=\frac{1}{\omega }\int d^{3}x\left[ \frac{1}{2}|D\psi |^{2}+%
\frac{1}{2}|\partial _{z}\psi |^{2}-\frac{1-t}{2}|\psi |^{2}+\frac{1}{2}%
|\psi |^{4}\right] ,
\end{equation}
by $x\rightarrow \xi x,y\rightarrow \xi y,z\rightarrow \frac{\xi z}{\gamma
^{1/2}},\psi ^{2}\rightarrow \frac{2\alpha T_{c}}{b^{\prime }}\psi ^{2},$
where $\gamma \equiv m_{c}/m_{ab}$ is anisotropy. The Ginzburg number is now
given by 
\begin{equation}
Gi\equiv \frac{1}{2}\left( \frac{32\pi e^{2}\kappa ^{2}\xi T_{c}\gamma ^{1/2}%
}{c^{2}h^{2}}\right) ^{2}
\end{equation}
Within the LLL approximation: 
\begin{equation}
f=\frac{F}{T}=\frac{1}{\omega }\int d^{3}x\left[ \frac{1}{2}|\partial
_{z}\psi |^{2}-\frac{1-t-b}{2}|\psi |^{2}+\frac{1}{2}|\psi |^{4}\right] .
\end{equation}
It also possesses an LLL scaling different from the 2D one. After a
rescaling $x\rightarrow x/\sqrt{b},y\rightarrow y/\sqrt{b},z\rightarrow
z\left( \frac{b\omega }{4\pi \sqrt{2}}\right) ^{-1/3},\psi ^{2}\rightarrow
\left( \frac{b\omega }{4\pi \sqrt{2}}\right) ^{2/3}\psi ^{2}$, the
dimensionless free energy becomes: 
\begin{equation}
f=\frac{1}{4\pi \sqrt{2}}\int d^{3}x\left[ \frac{1}{2}|\partial _{z}\psi
|^{2}+a_{T}|\psi |^{2}+\frac{1}{2}|\psi |^{4}\right] .
\end{equation}
The 3D reduced temperature is: 
\begin{mathletters}
\begin{equation}
a_{T}=-\left( \frac{b\omega }{4\pi \sqrt{2}}\right) ^{-2/3}\frac{1-t-b}{2}.
\label{aT3D}
\end{equation}
The relation between the original and scaled quantity (the 3D Jacobian
contains an ultraviolet divergent term which will be will cancel with
corresponding one loop divergence and is not written here) is: 
\end{mathletters}
\begin{eqnarray}
-\frac{T\log Z}{V} &=&\frac{T}{4\pi \sqrt{2}}\frac{V^{^{\prime }}}{V}\frac{%
\left( -4\pi \sqrt{2}\log Z_{r}J\right) }{V^{^{\prime }}}  \nonumber \\
&=&\frac{T}{4\pi }\frac{\sqrt{\gamma }b}{\xi ^{3}}\left( \frac{b\omega }{%
4\pi \sqrt{2}}\right) ^{1/3}f_{eff}.
\end{eqnarray}

\subsection{General description of the optimized gaussian perturbation
theory for scalar fields.}

We will use a variant of OPT, the optimized gaussian series \cite{Kleinert}
to study the vortex liquid. It is based on the ''principle of minimal
sensitivity'' \ idea \cite{Stevenson}, first introduced in quantum
mechanics. Any perturbation theory starts from dividing the Hamiltonian into
a solvable ''large'' part and a perturbation. Since we can solve any
quadratic Hamiltonian we have a freedom to choose ''the best'' such
quadratic part. Quite generally such an optimization converts an asymptotic
series into a convergent one (see a comprehensive discussion, references and
a proof in \cite{Kleinert}). Here we describe the implementation of the OPT
idea using a simple model of a real scalar fields $\phi $%
\begin{equation}
f=-\frac{1}{2}\phi D^{-1}\phi +V(\phi ).  \label{phi4}
\end{equation}
The free energy is divided into the ''large'' quadratic part and a
perturbation introducing variational parameter function $G^{-1}:$ 
\begin{equation}
f=K+\alpha v,\;K=\frac{1}{2}\phi G^{-1}\phi ,\;\;v=f-\frac{1}{2}\phi
G^{-1}\phi .  \nonumber
\end{equation}
Here the auxiliary parameter $\alpha $ was introduced to generate a
perturbation theory. It will be set to one at the end of calculation.
Expanding the logarithm of the statistical sum to order $\alpha ^{n+1}$ 
\begin{eqnarray}
Z &=&\int {\cal D}\phi \exp (-K)\exp (-\alpha v)=\int {\cal D}\phi \sum_{i=0}%
\frac{1}{i!}\left( \alpha v\right) ^{i}\exp (-K), \\
\widetilde{f}_{n}[G] &=&-\log Z=-\log \left[ \int {\cal D}\phi \exp (-K)%
\right] -\sum_{i=1}^{n+1}\frac{\left( -\alpha \right) ^{i}}{i!}\left\langle
v^{i}\right\rangle _{K},  \nonumber
\end{eqnarray}
where $\left\langle {}\right\rangle _{K}$ denotes the sum of all the
connected Feynman diagrams with $G$ as a propagator and then taking $\alpha
\rightarrow 1,$ we obtain a functional of $G$. To define the $n^{th}$ order
OPT approximant $f_{n}$ one minimizes $\widetilde{f}_{n}[G]$ with respect to 
$G$: 
\begin{equation}
f_{n}=\min_{G}\widetilde{f}_{n}[G].  \label{min}
\end{equation}
Till now the method has been applied and comprehensively investigated in
quantum mechanics only (\cite{Kleinert} and references therein) although
attempts in field theory have been made \cite{Stevenson}.

\section{OPT in the Ginzburg - Landau model}

\subsection{2D}

Due to the translational symmetry of the vortex liquid there is only one
variational parameter, $\varepsilon ,$ in the free energy defined by: 
\begin{eqnarray}
K &=&\frac{\varepsilon }{4\pi }|\psi |^{2} \\
v &=&+\frac{\alpha }{4\pi }\left[ a_{H}|\psi |^{2}+\frac{1}{2}|\psi |^{4}%
\right]   \nonumber
\end{eqnarray}
where $a_{H}\equiv a_{T}-\varepsilon $. It is convenient to use the quasi -
momentum eigenfunctions similar to those used extensively in the vortex
lattice: 
\begin{equation}
\varphi _{{\bf k}}=\sqrt{\frac{2\pi }{\sqrt{\pi }a_{\bigtriangleup }}}%
\sum\limits_{l=-\infty }^{\infty }\exp \left\{ i\left[ \frac{\pi l(l-1)}{2}+%
\frac{2\pi (x-k_{y})}{a_{\bigtriangleup }}l-xk_{x}\right] -\frac{1}{2}%
(y+k_{x}-\frac{2\pi }{a_{\bigtriangleup }}l)^{2}\right\} ,
\end{equation}
where $a_{\bigtriangleup }=\sqrt{\frac{4\pi }{\sqrt{3}}}$. We expand 
\begin{equation}
\psi (x)=\int_{{\bf k}}\frac{\varphi _{{\bf k}}(x)}{\left( \sqrt{2\pi }%
\right) ^{2}}\psi (k)
\end{equation}
Then the propagator in the quasi - momentum basis is: 
\begin{equation}
\left\langle \psi (k)\psi (l)\right\rangle =\frac{4\pi }{\varepsilon }\delta
\left( k+l\right) 
\end{equation}
In the coordinate space: 
\begin{eqnarray}
\left\langle \psi ^{\ast }(x_{1},y_{1})\psi (x_{2},y_{2})\right\rangle  &=&%
\frac{2}{\varepsilon }\exp \left[ -\frac{i}{2}\left( x_{1}-x_{2}\right)
\left( y_{1}+y_{2}\right) \right] \times  \\
&&\exp \left\{ -\frac{1}{4}\left[ \left( x_{1}-x_{2}\right) ^{2}+\left(
y_{1}-y_{2}\right) ^{2}\right] \right\} .  \nonumber
\end{eqnarray}
The Feynman rules are given on Fig. 1. We have a propagator denoted by a
directed line, Fig.1a, connecting two points $(x_{1},y_{1})$ to $%
(x_{2},y_{2})$. For the first term in $v$, we have a vertex represented by a
dot on a line, Fig. 1c with a value of $\frac{\alpha }{4\pi }a_{H}$. The
second term is a four line vertex, Fig. 1b, with a value of $\frac{\alpha }{%
4\pi }\frac{1}{2}$. To calculate the effective energy density $\
f_{eff}=-4\pi \ln Z$, we draw all the connected vacuum diagrams. Then one of
the coordinates is fixed, and all the other are integrated out. We
calculated directly diagrams up to the three loop order shown on Fig.2,3,4
with the following result. 
\begin{eqnarray}
\widetilde{f}_{0} &=&2\ast (\frac{2}{\varepsilon ^{2}}+\frac{a_{H}}{%
\varepsilon }+\log \frac{\varepsilon }{4\pi ^{2}})  \nonumber \\
\widetilde{f}_{1} &=&\widetilde{f}_{0}-\frac{1}{\varepsilon ^{4}}\left(
18+8a_{H}\varepsilon +a_{H}^{2}\varepsilon ^{2}\right)  \\
\widetilde{f}_{2} &=&\widetilde{f}_{1}+\frac{2}{9\varepsilon ^{6}}\left(
662+324a_{H}\varepsilon +54a_{H}^{2}\varepsilon ^{2}+3a_{H}^{3}\varepsilon
^{3}\right)   \nonumber
\end{eqnarray}
However to take advantage of the existing long series of the non optimized
gaussian expansion, we found a relation of the OPE to these series.
Originally Thouless and Ruggeri calculated these series $f_{eff}$ to sixth
order, but it was subsequently extended to 12$^{th}$ by Hikami et al \cite
{Hikami} and to 13$^{th}$ by Hu and MacDonald \cite{Hu}. It can be presented
using variable $x$ introduced by Thouless and Ruggeri \cite{Ruggeri} 
\begin{equation}
x=\frac{1}{\varepsilon ^{2}},\varepsilon =\frac{1}{2}\left( a_{T}+\sqrt{%
a_{T}^{2}+16}\right) 
\end{equation}
as follows: 
\begin{equation}
f_{eff}=2\log \frac{\varepsilon }{4\pi ^{2}}+2f_{2D}\left( x\right)
,f_{2D}\left( x\right) =\sum_{n=1}^{\infty }c_{n}x^{n}.  \label{serie}
\end{equation}
The coefficients are given in Table 1.

\begin{center}
\ \ \ {\bf Table 1.}

Coefficient $c$ and $z$ in 2D.

\begin{tabular}{|c|c|c|}
\hline
$n$ & $c_{n}$ & $z_{n-1}$ \\ \hline
$1$ & $-2$ & $-4$ \\ \hline
$2$ & $-1$ & $-6$ \\ \hline
$3$ & $\frac{38}{9}$ & $-12.239721181139888$ \\ \hline
$4$ & $-39-\frac{29}{30}$ & $-7.508888400035477$ \\ \hline
$5$ & $471.39659451659446$ & $-7.349933383279474$ \\ \hline
$6$ & $-6471.5625749551446$ & $-14.152646217045422$ \\ \hline
$7$ & $101279.32784597063$ & $-9.961364397930787$ \\ \hline
$8$ & $-1779798.7875947522$ & $-9.174960576928443$ \\ \hline
$9$ & $34709019.614363678$ & $-15.232548389083844$ \\ \hline
$10$ & $-744093435.66822231$ & $-11.629924499110746$ \\ \hline
$11$ & $17399454123.559521$ & $-10.8399817525306$ \\ \hline
$12$ & $-440863989257.28510$ & $-15.9366927661989$ \\ \hline
$13$ & $12035432945204.531$ & $-12.753308785106007$ \\ \hline
\end{tabular}
\end{center}

We can obtain all the OPT diagrams which do not appear in the gaussian
theory by insertions of bubbles and vertex Fig1c.insertions from the
diagrams contributing to the nonoptimized theory. Bubbles or ''cacti''
diagrams, see Fig.5 are effectively inserted in eq.(\ref{serie}) by
technique known in field theory \cite{Barnes}: 
\begin{eqnarray}
\ f_{eff} &=&2\log \frac{\varepsilon _{1}}{4\pi ^{2}}+2f_{2D}\left( x\right)
;  \nonumber \\
x &=&\frac{\alpha }{\varepsilon _{1}^{2}},\varepsilon _{1}=\frac{1}{2}\left(
\varepsilon _{2}+\sqrt{\varepsilon _{2}^{2}+16\alpha }\right) .
\end{eqnarray}
Summing up all the insertions of the mass vertex is achieved by

\begin{equation}
\varepsilon _{2}=\varepsilon +\alpha a_{H}.
\end{equation}
\ We then expand $f_{eff}$ to order $\alpha ^{n+1},$ and then taking $\alpha
=1$, to obtain $f$ $_{n}$.Calculating $f$ $_{n}$ that way, we checked that
indeed the first three orders agree with the calculation performed by a
direct calculation. Here a few more terms are displayed: 
\begin{eqnarray}
\widetilde{f}_{3} &=&\widetilde{f}_{2}-\frac{8133}{5\varepsilon ^{8}}-\frac{%
2648a_{H}}{3\varepsilon ^{7}}-\frac{180a_{H}^{2}}{\varepsilon ^{6}}-\frac{%
16a_{H}^{3}}{\varepsilon ^{5}}-\frac{a_{H}^{4}}{2\varepsilon ^{4}}  \nonumber
\\
\widetilde{f}_{4} &=&\widetilde{f}_{3}+\frac{21894.3}{\varepsilon ^{10}}+%
\frac{13012.8a_{H}}{\varepsilon ^{9}}+\frac{3089.33a_{H}^{2}}{\varepsilon
^{8}}+ \\
&&\frac{360a_{H}^{3}}{\varepsilon ^{7}}+\frac{20a_{H}^{4}}{\varepsilon ^{6}}+%
\frac{0.4a_{H}^{5}}{\varepsilon ^{5}}  \nonumber
\end{eqnarray}
The nth OPT approximant $f_{n}$ \ is obtained by minimization of $\widetilde{%
f}_{n}(\varepsilon )$ with respect to $\varepsilon $: 
\begin{equation}
\left( \frac{\partial }{\partial \varepsilon }-\frac{\partial }{\partial
a_{H}}\right) \widetilde{f}_{n}\left( \varepsilon ,a_{H}\right) =0.
\label{mini}
\end{equation}
The above equation is equal to $\frac{1}{\varepsilon ^{2n+3}}$ times a
polynomial $g_{n}\left( z\right) $ of order $n$ in $z\equiv \varepsilon
\cdot a_{H}$. That eq.(\ref{mini}) is of this type can be seen by noting
that\ the function $f$ depends on combination $\frac{\alpha }{\left(
\varepsilon +\alpha a_{H}\right) ^{2}}$ only. We were unable to prove this
rigorously, but have checked it to the 40$^{th}$ order in $\alpha .$ This
property simplifies greatly the task: one has to find roots of polynomials
rather than solving transcendental equations. There are $n$ (real or
complex) solutions for $g_{n}\left( z\right) =0$. However (as in the case of
anharmonic oscillator \cite{Kleinert}) \ the best root is the real root with
the smallest absolute value,. The roots $z_{n}$ for $n=0$ to $n=12$ are
given in Table 1.

We then obtain $\varepsilon (a_{T})=\frac{a_{T}+\sqrt{a_{T}^{2}-4z_{n}}}{2}$
solving $z_{n}=\varepsilon \cdot a_{H}=\varepsilon a_{T}-\varepsilon ^{2}$.
For $z_{0}=-4,$we obtain the gaussian result, dashed line marked ''T0'' on
Fig. 6. \ 

\subsection{3D}

In the 3D, the LLL Ginzburg - Landau model, we set: 
\begin{eqnarray}
K &=&\frac{1}{4\pi \sqrt{2}}\left( \varepsilon |\psi |^{2}+\frac{1}{2}%
|\partial _{z}\psi |^{2}\right)   \nonumber \\
v &=&+\frac{\alpha }{4\pi \sqrt{2}}\left[ a_{H}|\psi |^{2}+\frac{1}{2}|\psi
|^{4}\right] 
\end{eqnarray}

and 
\begin{equation}
\psi (x)=\int_{k_{3}}\int_{{\bf k}}\frac{\exp \left[ izk_{3}\right] \varphi
_{{\bf k}}(x)}{\left( \sqrt{2\pi }\right) ^{3}}\psi (k)
\end{equation}
The propagator is 
\begin{equation}
\left\langle \psi (k)\psi (l)\right\rangle =\frac{4\pi \sqrt{2}}{\varepsilon
+\frac{k_{3}^{2}}{2}}\delta \left( k+l\right) 
\end{equation}
or in the coordinate space: 
\begin{eqnarray}
\left\langle \psi (x_{1},y_{1,}z_{1})\psi (x_{2},y_{2},z_{2})\right\rangle 
&=&\frac{\sqrt{2}}{\pi }\int_{k_{3}}\frac{\exp \left[ ik_{3}\left(
z_{1}-z_{2}\right) \right] }{\varepsilon +\frac{k_{3}^{2}}{2}}\exp \left[ -%
\frac{i}{2}\left( x_{1}-x_{2}\right) \left( y_{1}+y_{2}\right) \right]
\times   \nonumber \\
&&\exp \left\{ -\frac{1}{4}\left[ \left( x_{1}-x_{2}\right) ^{2}+\left(
y_{1}-y_{2}\right) ^{2}\right] \right\} .
\end{eqnarray}
Thus the propagator in the coordinate space factorizes into a function of
coordinates $(x,y)$ perpendicular to magnetic field and a function of the
coordinate $z$ parallel to it. The mass insertion vertex, Fig. 1c, now has a
value of $\frac{\alpha }{4\pi \sqrt{2}}a_{H}$, while the four line vertex is 
$\frac{\alpha }{8\pi \sqrt{2}}$. The calculation is basically the same as in
2D, the only difference being extra integrations over $k_{3}$. However since
the propagator factorizes,\ these integrations can be reduced to
corresponding integrations in quantum mechanics of the anharmonic oscillator 
\cite{Ruggeri,Okopinska}.

Again we can take an advantage of existing long series of the nonoptimized
gaussian expansion \cite{Ruggeri,Hikami}. The results to seventh order are: 
\begin{eqnarray}
f_{eff} &=&4\sqrt{\varepsilon }+4\sqrt{\varepsilon }f_{3d}\left( x\right) ; 
\nonumber \\
f_{3D}\left( x\right)  &=&\sum c_{n}x^{n},\;x=\frac{1}{\sqrt{\varepsilon ^{3}%
}},
\end{eqnarray}
where $\sqrt{E}$ is given by a solution of the cubic gap equation $\left( 
\sqrt{\varepsilon }\right) ^{3}-a_{T}\sqrt{\varepsilon }-4=0$: 
\begin{equation}
\sqrt{\varepsilon }=a_{T}\left( 54+3\sqrt{324-3a_{T}^{3}}\right) ^{-1/3}+%
\frac{1}{3}\left( 54+3\sqrt{324-3a_{T}^{3}}\right) ^{1/3}
\end{equation}
and coefficient $c_{n}$ are listed in Table 2:\ \ \ 

\begin{center}
{\bf Table 2.}

Coefficients $c$ and $z$ in 3D

\begin{tabular}{|c|c|c|}
\hline
$n$ & $c_{n}$ & $z_{n-1}$ \\ \hline
$1$ & $-2$ & $-4$ \\ \hline
$2$ & $-0.5$ & $-5$ \\ \hline
$3$ & $1.583333333$ & $-8.80317864821579$ \\ \hline
$4$ & $-12.667361111$ & $-6.187603657880674$ \\ \hline
$5$ & $125.59552619$ & $-5.960012621607176$ \\ \hline
$6$ & $-1430.5928959$ & $-9.47212746817198$ \\ \hline
$7$ & $18342.765997$ & -$7.430474107869646$ \\ \hline
$8$ & $-261118.67703$ & $-6.907260317913621$ \\ \hline
$9$ & $4084812.307$ & $-9.8195351835546$ \\ \hline
\end{tabular}
\end{center}

Similarly the OPT formula for the effective energy density can be obtained
by using the generational function: 
\begin{equation}
f_{eff}=4\sqrt{\varepsilon _{1}}+4\sqrt{\varepsilon _{1}}f_{3D}\left(
x\right) ,\;x=\frac{\alpha }{2\left( \sqrt{\varepsilon _{1}}\right) ^{3}}
\end{equation}
and $\sqrt{\varepsilon _{1}}$ is given by a solution of equation 
\begin{equation}
\left( \sqrt{\varepsilon _{1}}\right) ^{3}-\varepsilon _{2}\sqrt{\varepsilon
_{1}}-4\alpha =0  \label{alphaex}
\end{equation}
with $\varepsilon _{2}=\varepsilon +\alpha a_{H}$. The solution of eq.(\ref
{alphaex}) can be obtained perturbatively in $\alpha $: 
\begin{eqnarray}
&&\sqrt{\varepsilon _{1}}=\sqrt{\varepsilon _{2}}+\frac{2\alpha }{%
\varepsilon _{2}}-\frac{6\alpha ^{2}}{\varepsilon _{2}^{5/2}}+\frac{32\alpha
^{3}}{\varepsilon _{2}^{4}}-\frac{210\alpha ^{4}}{\varepsilon _{2}^{11/2}}+%
\frac{1536\alpha ^{5}}{\varepsilon _{2}^{7}}  \nonumber \\
&&-\frac{12012\alpha ^{6}}{\varepsilon _{2}^{17/2}}+\frac{98304\alpha ^{7}}{%
\varepsilon _{2}^{10}}-\frac{831402\alpha ^{8}}{\varepsilon _{2}^{23/2}}+%
\frac{7208960\alpha ^{9}}{\varepsilon _{2}^{13}}+...
\end{eqnarray}
Expanding $f_{eff}$ \ in $\alpha $ to order $n+1$, then one then sets $%
\alpha =1$ to obtain $\widetilde{f}_{n}$.

We list here first few OPT approximants $\widetilde{f}_{n}$ 
\begin{eqnarray}
\widetilde{f}_{0} &=&4\sqrt{\varepsilon }+\frac{2a_{H}}{\sqrt{\varepsilon }}+%
\frac{4}{\varepsilon }  \nonumber \\
\widetilde{f}_{1} &=&\widetilde{f_{0}}-\frac{1}{2\sqrt{\varepsilon ^{5}}}%
\left( 17+8a_{H}\sqrt{\varepsilon }+a_{H}^{2}\varepsilon \right)   \nonumber
\\
\widetilde{f_{2}} &=&\widetilde{f_{1}}+\frac{1}{24\varepsilon ^{4}}\left(
907+510a_{H}\sqrt{\varepsilon }+96a_{H}^{2}\varepsilon +6a_{H}^{3}\sqrt{%
\varepsilon ^{3}}\right)  \\
\widetilde{f_{3}} &=&\widetilde{f_{2}}-\frac{228.8335069417501}{\sqrt{%
\varepsilon ^{11}}}-\frac{151.166666666a_{H}}{\varepsilon ^{5}}  \nonumber \\
&&-\frac{37.1875a_{H}^{2}}{\sqrt{\varepsilon ^{11}}}-\frac{4a_{H}^{3}}{%
\varepsilon ^{4}}-\frac{0.15625a_{H}^{4}}{\sqrt{\varepsilon ^{7}}}  \nonumber
\end{eqnarray}
The OPT $n^{th}$ order result $f_{n}(a_{T})$ is obtained optimizing $%
\widetilde{f}_{n}$ by varying $\varepsilon :$ 
\begin{equation}
\left( \frac{\partial }{\partial \varepsilon }-\frac{\partial }{\partial
a_{H}}\right) \widetilde{f}_{n}\left( \varepsilon ,a_{H}\right) =0.
\label{mini2}
\end{equation}
Similarly to eq.(\ref{mini}) in 2D this is equal to $\frac{1}{\varepsilon ^{%
\frac{3n}{2}+2}}g_{n}\left( z\right) ,$ where now $z\equiv a_{H}\sqrt{%
\varepsilon }$ and $g_{n}\left( z\right) $ is a rank $n$ polynomial. Solving 
$g_{n}\left( z\right) $ and choosing a real root with the smallest absolute
value \cite{Kleinert}, we obtain $z_{n}$ listed in table 2 up to $n=8$. Then
we solve for $\sqrt{\varepsilon }$ the equation $z=a_{H}\sqrt{\varepsilon }%
=\left( a_{T}-\varepsilon \right) \sqrt{\varepsilon }$. The solution is 
\begin{equation}
\sqrt{\varepsilon }=2^{1/3}a_{T}\left( -27z+\sqrt{-108a_{T}^{3}+729z^{2}}%
\right) ^{-1/3}+\frac{1}{32^{1/3}}\left( -27z+\sqrt{-108a_{T}^{3}+729z^{2}}%
\right) ^{1/3}
\end{equation}

\section{Results and comparison with other theories and experiments}

\subsection{Energy, precision of OPT}

On Fig. 6 we present OPT for orders $n=0$ (gaussian)$,1,3,4,5,6,8,9,12$
together with several orders ($T0,...12)$ of the nonoptimized high
temperature expansion in 2D . The values of free energy of 2D \ and 3D
models for several $a_{T}$ are tabulated in Table 3 and Table 4
respectively. One clearly observes that in 2D the OPT series converge above $%
a_{T}=-2.5$ and diverge below $a_{T}=-3.5.$ On the other hand, the non
optimized series never converge despite the fact that above $a_{T}=2$ first
few approximants provide a quite precise estimate consistent with OPT. Above 
$a_{T}=4$ the liquid becomes essentially a normal metal and fluctuations
effects are negligible (see Fig. 7, 8) and are hard to measure. Therefore
the information the OPT provides is essential to compare with experiments on
magnetization and specific heat.

\begin{center}
{\bf Table 3.}

Free energy $f_{n}$ at different orders (modula constant $-2\log 4\pi ^{2}$)

\begin{tabular}{|c|c|c|c|c|}
\hline
$a_{T}$ & $-2$ & $-1.5$ & $-1$ & $-0.5$ \\ \hline
$f_{0}$ & $-2.19416$ & $-1.42941$ & $-0.749027$ & $-0.146255$ \\ \hline
$f_{1}$ & $-2.77516$ & $-1.80556$ & $-0.988706$ & $-0.297222$ \\ \hline
$f_{3}$ & $-2.53854$ & $-1.68294$ & $-0.925643$ & $-0.264857$ \\ \hline
$f_{4}$ & $-2.55889$ & $-1.69143$ & $-0.92912$ & $-0.266258$ \\ \hline
$f_{6}$ & $-2.70076$ & $-1.74015$ & $-0.945544$ & $-0.271734$ \\ \hline
$f_{7}$ & $-2.62447$ & $-1.71822$ & $-0.939384$ & $-0.270031$ \\ \hline
$f_{9}$ & $-2.51533$ & $-1.6923$ & $-0.933365$ & $-0.268653$ \\ \hline
$f_{10}$ & $-2.59943$ & $-1.70944$ & $-0.936772$ & $-0.269318$ \\ \hline
$f_{12}$ & $-2.72613$ & $-1.73113$ & $-0.940395$ & $-0.269915$ \\ \hline
\end{tabular}
\end{center}

If precision is defined as $\left( f_{12}-f_{10}\right) /f_{10}$, we obtain $%
4.87\%,1.27\%,0.387\%,0.222\%,$ \ $0.032\%$ at $a_{T}=-2,-1.5,-1,-0.5,0$
respectively. We chose approximants $n=0$ $,1,3,4,6,7,9,10,12$ \ because
they are ''the best roots'' in a sense defined in ref. \cite{Kleinert},
section 5. For comparison with other theories and experiments on Fig. 7 and
8 we use the 10$^{th}$ approximant.

In 3D the picture is much the same. The series converge above $a_{T}=-4.5$
and diverge below $a_{T}=-5.5.$ The non optimized series are useful only
above $a_{T}=-1$.

\begin{center}
{\bf Table 4.}

Free energy $f_{n}$ at different orders for 3D

\begin{tabular}{|l|l|l|l|l|}
\hline
$a_{T}$ & $-5$ & $-3$ & $-1.5$ & $-1$ \\ \hline
$f_{0}$ & $-4.73313$ & $\ \ \ \ \;0$ & $2.65763$ & $3.41112$ \\ \hline
$f_{1}$ & $-6.493$ & $-0.375697$ & $2.53901$ & $3.32829$ \\ \hline
$f_{2}$ & $-6.92585$ & $-0.427383$ & $2.5287$ & $3.3222$ \\ \hline
$f_{3}$ & $-5.27595$ & $-0.280923$ & $2.55551$ & $3.338$ \\ \hline
$f_{4}$ & $-5.68059$ & $-0.292936$ & $2.55455$ & $3.33757$ \\ \hline
$f_{5}$ & $-4.68076$ & $-0.265834$ & $2.55647$ & $3.33839$ \\ \hline
$f_{6}$ & $-7.32654$ & $-0.313048$ & $2.55364$ & $3.33722$ \\ \hline
$f_{7}$ & $-5.33149$ & $-0.301797$ & $2.55392$ & $3.33731$ \\ \hline
$f_{8}$ & $-8.01907$ & $-0.316175$ & $2.55359$ & $3.3372$ \\ \hline
\end{tabular}
\end{center}

We define the precision as $\left( f_{7}-f_{4}\right) /f_{7}$. $f_{4}$ and $%
f_{7}$ are the best roots among the sequences. Then we obtain $\ \ \
6.\,55\%,2.\,\allowbreak 94\%,0.0247\%,0.00779222\%$, at $%
a_{T}=-5,-3,-1.5,-1 $ respectively.

\subsection{Other theories}

We compare with other theoretical treatments of the same model. A direct
method is the Monte Carlo simulation of the same model. The 2D model was
simulated by Moore, Kato and Nagaosa , and Hu McDonald. The circles on Fig.
8 for specific heat are the results of the Monte Carlo simulation of the LLL
system by Kato and Nagaosa in ref. \cite{MC} performed with 256 vortices. In
3D the model was simulated with 100 vortices by Sasik and Stroud \cite{Sasik}%
, magnetization data are compared with our results on Fig.9.

An analytic theory used successfully to fit the magnetization and the
specific heat data \cite{Pierson} was developed in \cite{Tesanovic}. Their \
free energy density is: 
\begin{eqnarray}
f_{eff} &=&-\frac{a_{T}^{2}U^{2}}{4}+\frac{a_{T}U}{2}\sqrt{\frac{%
U^{2}a_{T}^{2}}{4}+2}+2arc\sinh \left[ \frac{a_{T}U}{2\sqrt{2}}\right]
\label{teseq} \\
U &=&\frac{1}{2}\left[ \frac{1}{\sqrt{2}}+\frac{1}{\sqrt{\beta _{A}}}+\tanh %
\left[ \frac{a_{T}}{4\sqrt{2}}+\frac{1}{2}\right] \left( \frac{1}{\sqrt{2}}-%
\frac{1}{\sqrt{\beta _{A}}}\right) \right] .  \nonumber
\end{eqnarray}
The corresponding magnetization and specific heat are shown as a dashed
lines on Fig.7 and 8 respectively.. The theory applies not only to the
liquid phase, but also to the solid although the transition is not seen
(should be considered as a 2\% effect not determined by the theory). At
large positive $a_{T}$ neglecting the exponentially small contributions to $%
U $, one obtains: 
\begin{eqnarray}
f_{eff} &=&-\frac{a_{T}^{2}}{8}+\frac{a_{T}}{2\sqrt{2}}\sqrt{\frac{a_{T}^{2}%
}{8}+2}+2arc\sinh \left[ \frac{a_{T}}{4}\right]  \nonumber \\
&=&1-2\log 2+2\log a_{T}+\frac{4}{a_{T}^{2}}-\frac{16}{a_{T}^{4}}+\frac{320}{%
3a_{T}^{6}}
\end{eqnarray}
On the other hand, \ the high temperature expansion of the optimized
gaussian is 
\begin{equation}
-2\log 4\pi ^{2}+2\log a_{T}+\frac{4}{a_{T}^{2}}-\frac{18}{a_{T}^{4}}+\frac{%
1324}{9a_{T}^{6}}
\end{equation}
One observes that the high temperature expansion of two theories are in
remarkable agreement up to the order $\frac{1}{a_{T}^{4}}$.

\subsection{Magnetization, 2D}

Experiments on great variety of layered high $Tc$ cuprates ($Bi$ or $Tl$ 
\cite{Magn2D} based) show that in 2D, magnetization curves for different
applied field intersect at a single point $(M^{\ast },T^{\ast })$.The range
of magnetic fields is surprisingly large (from several hundred $Oe$ to
several $Tesla$). Assuming this it is easy to derive the scaled LLL
magnetization just from the existence of the point. The dimensionless LLL
magnetization is defined as \cite{Solid} 
\begin{equation}
m(a_{T})=-\frac{df_{eff}(a_{T})}{da_{T}}
\end{equation}
and the measure magnetization is 
\begin{equation}
M=-\frac{e^{\ast }{h}}{m_{ab}}\left\langle |\psi |^{2}\right\rangle =-\frac{%
e^{\ast }{h}}{m_{ab}}|\psi _{r}|^{2}\sqrt{\frac{b\omega }{4\pi }},
\end{equation}
where $\psi $ is the order parameter of the original model, and $\psi _{r}$
is the rescaled one, which is equal to $\frac{df_{eff}(a_{T})}{da_{T}}$.
Thus 
\begin{equation}
M=\frac{e^{\ast }{h}}{m_{ab}}\sqrt{\frac{b\omega }{4\pi }}m(a_{T}).
\label{magde}
\end{equation}
Using the definition of \ $a_{T}=-\eta \frac{1-t-b}{\sqrt{bt}},\eta =\left(
2\pi ^{2}Gi\right) ^{-1/4},b$ can be written \ as 
\begin{equation}
b=t\left( \frac{a_{T}}{2\eta }\pm \sqrt{\frac{1-t}{t}+\frac{a_{T}^{2}}{4\eta
^{2}}}\right) ^{2}.
\end{equation}
Thus eq.(\ref{magde}) implies that

\begin{eqnarray}
m\left( a_{T}\right) &=&\frac{m_{ab}M}{eh}\frac{\eta }{\sqrt{bt}}  \nonumber
\\
&=&\frac{\eta m_{ab}M}{eht}\times \frac{1}{\left| \sqrt{\frac{1-t}{t}+\frac{%
a_{T}^{2}}{4\eta ^{2}}}\pm \frac{a_{T}}{2\eta }\right| } \\
&=&\frac{m_{ab}M\eta }{eh\left| 1-t\right| }\left| \sqrt{\frac{1-t}{t}+\frac{%
a_{T}^{2}}{4\eta ^{2}}}\pm \frac{a_{T}}{2\eta }\right|  \nonumber
\end{eqnarray}
If we assume that the experimental observation that all the magnetization
curves intersect at some point $(T^{\ast },M^{\ast })$, $m\left(
a_{T}\right) $ is 
\begin{eqnarray}
m\left( a_{T}\right) &=&C_{1}\left( a_{T}\pm \sqrt{C_{2}+a_{T}^{2}}\right) ;
\label{fixmag} \\
C_{1} &=&\frac{m_{ab}M^{\ast }}{e^{\ast }h\left| 1-t^{\ast }\right| }%
,\;C_{2}=4\eta ^{2}\frac{1-t^{\ast }}{t^{\ast }}.  \nonumber
\end{eqnarray}
On the other hand, if we require that the first two terms of the high
temperature expansion of eq.(\ref{fixmag}) and the high temperature
expansion of the magnetization are equal, one finds that. 
\[
C_{1}=\frac{1}{4}\ ;C_{2}=16 
\]
When we plot this line on Fig.7 (the dotted line) we find that at lower
temperatures the magnetization is overestimated. On the other hand
magnetization of the theory of Tesanovic et al (the dashed line on Fig. 7)
underestimate the magnetization. The OPE results are consistent with the
data within the precision range till the radius of convergence $a_{T}=-3.$
It is important to note that deviations of both the phenomenological formula
eq.(\ref{fixmag}) and the Tesanovic's are clearly beyond our precision range.

We conclude therefore that although the theory of Tesanovic et al is very
good at high temperatures (deviations only at the order $1/a_{T}^{4}$) they
become of the order 5-10\% at $a_{T}=-3$. The advantage of this theory is
however that it interpolated smoothly to the solid and never deviates more
than 10\%. The coincidence of the intersection of all the lines at the same
point $(T^{\ast },M^{\ast })$ cannot be exact. Like in 3D it is just
approximate, although the approximation is quite good especially at high
magnetic fields.

\subsection{Specific heat, 2D}

Specific heat OPE result is compared on Fig. 8 with Monte Carlo simulation
of the same model by Kato and Nagaosa \cite{MC} (black circles), the
phenomenological formula following from eq.(\ref{fixmag}) (dotted line) and
the theory of Tesanovic et al \cite{Tesanovic} (the dashed line). The
agreement with the direct MC simulation is very good.

\subsection{Magnetization in 3D}

We compare here our results on the LLL scaled magnetization with the Monte
Carlo simulation of the LLL system by Sasik and Stroud \cite{Sasik}. They
are actually more precise in 3D. The Fig. 9 contains several OPE
approximants ($n=0,1,2,3,4,7,8$) and their data on all three magnetic fields
(representing $2T$, $3T$ and $5T$ in model YBCO). According to the criterion
of the ''best root'' the best approximant should be $n=7$. Clearly up to the
radius of convergence the agreement is within the expected precision.

\section{Conclusion}

In this paper we obtained the optimized perturbation theory results for the
LLL Ginzburg - Landau model.

to both the 2D and 3D LLL model. It allows to obtain a convergent series
(rather than asymptotic).The magnetization and specific heat of vortex
liquids with definite precision are calculate. One the basis of this one can
make several definitive qualitative conclusions. The intersection of the
magnetization lines in only approximate not only in 3D (the result already
observed in Monte Carlo simulation \cite{Sasik}), but also in 2D. The theory
by Tesanovic \cite{Tesanovic}, which uses completely different ideas,
describes the physics remarkably well in high temperatures and deviates on
the 5-10\% precision level at $a_{T}=-2$ in 2D .

\acknowledgments
We are grateful to our colleagues A. Knigavko and T.K. Lee for numerous
discussions and encouragement and Z. Tesanovic for explaining his work to
one of us and sharing his insight. The work was supported by NSC of Taiwan
grant NSC..

\newpage

\begin{center}
{\Huge Figure captions}
\end{center}


{\LARGE Fig. 1}

Feynman rules. Fig.1a, Fig.1b,Fig.1c are propagator, four line vertex, mass
insertion vertex.

{\LARGE Fig. 2}

Feynmann diagrams for $\widetilde{f}_{n}[G]$ for $n=0$.

{\LARGE Fig. 3}

Fig.3 and Fig.2 are Feynmann diagrams for $\widetilde{f}_{n}[G]$ for $n=1$.

{\LARGE Fig. 4}

Fig4, Fig.3 and Fig.2 are Feynmann diagrams for $\widetilde{f}_{n}[G]$ for $%
n=2$.

{\LARGE Fig. 5}

Summing all bubble diagrams.

{\LARGE Fig. 6}

2D OPT energy and non-optimized energy at different orders (denoted by
numbers and T plus numbers respectively). One can see clearly OPT series are
convergent, for example at $-2$.

{\LARGE Fig. 7}

2D Magnetization plot: experimental data (Jin. {\it et.al} in ref.\cite
{Magn2D}), magnetization from OPT for $n=10$ (line) , from phenomenological
formula (dotted line) and Tesanovic' theory (dash-dotted line)

{\LARGE Fig. 8}

2D specific heat plot. Monte Carlo data by Kato and Nagaosa in ref.\cite{MC}%
, specific heat from OPT for $n=10$ (line) , from phenomenological formula
(dotted line) and Tesanovic' theory (dash-dotted line)

{\LARGE Fig. 9}

3D Magnetization plot. Monte Carlo data by Sasik and Stroud in ref.\cite{MC}%
, specific heat from OPT of different orders denoted by numbers. $n=4,7$ \
are best approximants among them.

\end{document}